\documentclass[letterpaper, 10 pt, conference]{ieeeconf}
    
    \IEEEoverridecommandlockouts 
    \makeatletter

    \let\proof\@undefined
    \let\endproof\@undefined
    \makeatother
    \let\labelindent\relax

    \usepackage{jhoagg}
    \usepackage{mathtools}
    \usepackage{amsfonts}
    \usepackage{amssymb,latexsym}
    \usepackage[psamsfonts]{eucal}
    \usepackage{amsthm}
    \usepackage{graphicx}
    \usepackage[inline]{enumitem}
    \usepackage{indentfirst}
    \usepackage{setspace}
    \usepackage{microtype}
    \usepackage{threeparttable}
    \usepackage{float}
    \usepackage{cleveref}
    \usepackage{afterpage}
    \usepackage{placeins}
    \usepackage{cancel}
    \usepackage{leftidx}
    \usepackage{lipsum}
    \usepackage{multicol}
    \usepackage{breqn}
    \usepackage[export]{adjustbox}
    \usepackage{comment}
    \usepackage[ruled, linesnumbered, lined]{algorithm2e}
    \usepackage[dvipsnames]{xcolor}
    \usepackage{xcolor}
    \usepackage[most]{tcolorbox}
    \usepackage{soul}
    \usepackage[noadjust]{cite}
    \usepackage{standalone}
    \usepackage{textgreek}

    \usepackage{cleveref}

    \usepackage{tikz}
    \usetikzlibrary{shapes,arrows,positioning,fit}

    \usepackage{pgf}
    \usepackage{pgfplots}
    \usepgflibrary{plothandlers}
    \usepgfplotslibrary{external} 
    \pgfkeys{/pgf/number format/.cd,1000 sep={}}  % Gets rid of comma when displaying 1,000
    \usetikzlibrary{calc,shapes.misc,plotmarks}
    \pgfplotsset{compat=1.13}
    
    \let\originalleft\left
    \let\originalright\right
    \renewcommand{\left}{\mathopen{}\mathclose\bgroup\originalleft}
    \renewcommand{\right}{\aftergroup\egroup\originalright}
    
    \usepackage{fancyhdr}
    \fancypagestyle{firststyle}
    {\fancyfoot{} 
    \fancyfoot[L]{\textit{Preprint submitted to 2024 American Control Conference (ACC)}}
     
    }
    
    \newcounter{thm} 
    \theoremstyle{definition}
    \newtheorem{theorem}[thm]{\indent Theorem}
    
    \newtheorem{assumption}{\indent Assumption}
    
    \theoremstyle{definition}
    \newtheorem{proposition}{\indent Proposition}    
    
    \newtheorem{lemma}{\indent Lemma}
    
    \newtheorem{remark}{\indent Remark}
    
    \newtheorem{corollary}{\indent Corollary}
    
    \newtheorem{definition}{\indent Definition}
    
    \newtheorem{example}{\indent Example}
    
    \newtheorem{fact}{\indent Fact}
    
    \newtheorem{conjecture}{\indent Conjecture}
    
    \newtheorem{experiment}{\indent Experiment}

    \allowdisplaybreaks

    \newlist{enumA}{enumerate}{1}
    \setlist[enumA,1]{label=(A\arabic*),leftmargin=1cm}

    \newlist{enumC}{enumerate}{1}
    \setlist[enumC,1]{label=(C\arabic*),leftmargin=1cm}

    \usepackage{cite}
    \usepackage{accents}
    
    \newlength\figureheight 
    \newlength\figurewidth

    \newcommand{\norm}[1]{\left\lVert#1\right\rVert}

    \DeclareMathOperator{\sgn}{sgn}

    \allowdisplaybreaks
    
    \graphicspath{ {Figures/} }
    \DeclareMathAlphabet{\mathcal}{OMS}{cmsy}{m}{n} % Use the standard calligraphy font

    \crefname{equation}{}{}
    \usepackage[ruled, linesnumbered, lined]{algorithm2e}
    \SetKwInput{KwInit}{Initialize}
    \SetKwFunction{SafeSet}{SafeSet}
    \SetKwFunction{GetSafe}{GetSafe}
    \SetKwFunction{GetUnsafe}{GetUnsafe}
    \SetCommentSty{mycommfont}
    \newcommand{\ubar}[1]{\underaccent{\bar}{#1}}
    \newlist{enumalph}{enumerate}{1}
    \setlist[enumalph]{label=\textit{(\alph*)}}

\begin{document}
\title{Electromagnetic Formation Flying with State and Input Constraints Using Alternating Magnetic Field Forces}
\author{Sumit S. Kamat, T. Michael Seigler, and Jesse B. Hoagg
% \thanks{P. Rabiee and J. B. Hoagg are with the Department of Mechanical and Aerospace Engineering, University of Kentucky, Lexington, KY, USA. (e-mail: pedram.rabiee@uky.edu, jesse.hoagg@uky.edu).}
% \thanks{This work is supported in part by the National Science Foundation (1849213, 1932105) and Air Force Office of Scientific Research (FA9550-20-1-0028).}
}
                                                                                                                                         
\maketitle

\begin{abstract}
This article presents a feedback control algorithm for electromagnetic formation flying with constraints on the satellites' states and control inputs. 
The algorithm combines several key techniques. 
First, we use alternating magnetic field forces to decouple the electromagnetic forces between each pair of satellites in the formation. 
Each satellite's electromagnetic actuation system is driven by a sum of amplitude-modulated sinusoids, where amplitudes are controlled in order to prescribe the time-averaged force between each pair of satellites.
Next, the desired time-averaged force is computed from a optimal control that satisfies state constraints (i.e., no collisions and an upper limit on intersatellite speeds) and input constraints (i.e., not exceeding satellite's apparent power capability). 
The optimal time-averaged force is computed using a single relaxed control barrier function that is obtained by composing multiple control barrier functions that are designed to enforce each state and input constraint. 
Finally, we demonstrate the satellite formation control method in a numerical simulation. 
\end{abstract}

\section{Introduction}\label{sec:introduction}

Electromagnetic formation flying (EMFF) is accomplished using electromagnetic coils onboard the satellites in a formation. 
Each satellite's electromagnetic coils generates a magnetic field, which interacts with the magnetic fields of the other satellites to create magnetic field forces.  
These magnetic field forces can be used to control the relative positions of satellites \cite{Kong2004,KWON2010,Porter2014}.
Control of 2 satellites with EMFF actuation is addressed in \cite{Elias2007} and demonstrated experimentally in \cite{Kwon2011}.

EMFF for more than 2 satellites is challenging because the intersatellite forces are nonlinear functions of the magnetic moments generated by all satellites in the formation as well as the relative positions of all satellites. 
In other words, there is complex coupling between the electromagnetic fields generated by all satellites in the formation, and this coupling ultimately leads to intersatellite forces.

EMFF for more than 2 satellites is addressed in \cite{Ahsun2006,Schweighart2010}.
However, these approaches require centralization of all measurement information.
A decentralized EMFF method is presented in \cite{Abbasi2022}. 
In this method, each satellite has access to measurements of its position and velocity relative to only its local neighbor satellites. 
In particular, \cite{Abbasi2022} addresses the complex intersatellite force coupling by using alternating magnetic field forces (AMFF). 
The key idea of AMFF is that a pair of alternating (e.g., sinusoidal) magnetic moments results in a nonzero average interaction force between the pair of satellites if and only if those alternating magnetic moments have the same frequency \cite{Youngquist2013,Nurge2016}.
Thus, \cite{Abbasi2022} uses a sum of frequency-multiplexed sinusoidal magnetic moments to achieve desired intersatellite forces between every pair of satellites. 
However, \cite{Abbasi2022} does not address state constraints (e.g., no satellite collisions) or input constraints (e.g., power limitations on the electromagnetic actuation system).

Control barrier functions (CBFs) can be used to determine controls that satisfy state constraints (e.g., \cite{wieland2007,ames2016,xiao2023,tan2021}). 
Specifically, CBFs are used to develop constraints that guarantee forward invariance of a set in which that the state constraints are satisfied.
CBFs are often implemented as constraints in real-time optimization control methods (e.g., quadratic programs) in order to guarantee state-constraint satisfaction while also minimizing a performance based cost \cite{ames2016}. 
Although CBFs are used to address state constraints rather than input constraints, \cite{rabiee2024b,rabiee2024a} presents a CBF-based approach that simultaneously addresses state constraints and input constraints. 
This approach uses control dynamics, which allows the input constraints to be transformed into controller-state constraints, and a log-sum-exponential soft-minimum to compose state-constraint CBFs and input-constraint CBFs into a single relaxed CBF.

This article presents a feedback control approach for EMFF using AMFF and a composite CBF in order to generate intersatellite forces that achieve formation and satisfy state and input constraints. 
We use the piecewise-sinusoidal magnetic moment approach in \cite{Abbasi2022} to prescribe the desired intersatellite forces between every pair of satellites; this approach is reviewed in \Cref{sec:Problem_Formulation}.
Then, \Cref{Section:Allocation} presents a new construction for the piecewise-sinusoidal amplitudes that achieve a prescribed intersatellite force. 
\Cref{sec:Formation control with state and input constraints} presents the main contribution of this article, namely, a feedback control for EMFF. 
We use model predictive control (MPC) to compute desired intersatellite forces, which achieve formation but may not respect state and input constraints. 
Then, we use the method from \cite{rabiee2024b,rabiee2024a} to construct a composite CBF that can be used to respect all state and input constraints. 
The desired intersatellite forces (from MPC) and the composite CBF are used together to construct a safe and optimal control.  
\Cref{sec:Formation Flying Simulation Results} demonstrates the EMFF approach in simulation.

% This article presents an approach to achieve EMFF with AMFF subject to satellite' state and input constraints. In section \ref{sec:Problem_Formulation} we present the translational dynamics of the system, followed by a review of the piecewise-sinusoidal magnetic moment approach in \cite{Abbasi2022} to prescribe the time-averaged intersatellite force. In section \ref{Section:Allocation} we present a construction of the piecewise-sinusoidal amplitudes.

% Section \ref{sec:Formation control with state and input constraints} presents the main contribution of this article. We determine a desired-time averaged force from an optimal control that satisfies state and input constraints. The optimal time-averaged force is determined using a single relaxed control barrier function that is obtained by composing multiple control barrier functions that are designed to enforce each state and input constraint. Finally, in section \ref{sec:Formation Flying Simulation Results} we demonstrate the satellite formation control method in a numerical simulation.

\section{Notation}

%Bold symbols are used to denote physical vectors. If $\mathbf{r}$ is a physical vector, then $| \mathbf{r} |$ is its magnitude. A frame $\mathcal{F}=\begin{bmatrix} \mathbf{i} &\mathbf{j} &\mathbf{k}\end{bmatrix}$ consists of a right-handed set of mutually orthogonal unit vectors $\mathbf{i}, \ \mathbf{j},$ and $\mathbf{k}$. If $x \in \mathbb{R}^{3}$, then $\norm{x}$ is it's 2 norm. If $x$ is a scalar, then $|x|$ is its absolute value.  

Let $\mathbb{N}$ denote the set of nonnegative integers.
We let $\mathcal I \triangleq \{1, ..., n\}$, where $n \in \mathbb{N}$ is the number of satellites in the formation, and $\mathcal P \triangleq \{(i,j) \in \mathcal I \times \mathcal I : i \neq j\}$, which is the set of ordered pairs.
Unless otherwise stated, statements in this paper that involve the subscript $i$ are for all $i \in \mathcal I$, and statements that involve the subscript $ij$ are for all $(i,j) \in \mathcal{P}$.

% For any $a \in \mathbb{R}^3$ where $a= \begin{bmatrix} a_1 &a_2 &a_3 \end{bmatrix}^{\mathrm{T}}$, we define
% \begin{align}
%     [ a ]_{\times} \triangleq \begin{bmatrix}
%         0 &-a_3 &a_2\\
%         a_3 &0 &-a_1\\
%         -a_2 &a_1 &0
%     \end{bmatrix}. \notag
% \end{align}

Let $\xi:\BBR^n \to \BBR$ be continuously differentiable. 
Then, $\xi^\prime :\BBR^n \to \BBR^{1 \times n}$ is defined as $\xi^\prime(x) \triangleq \pderiv{\xi(x)}{x}$. 
The Lie derivative of $\xi$ along the vector fields of $g:\BBR^n \to \BBR^{n \times m}$ is $L_g \xi(x) \triangleq \xi^\prime(x) g(x)$.
The boundary of the set $\SA \subseteq \BBR^n$ is denoted by $\mbox{bd }\SA$.

% Throughout this paper, we assume that all functions are sufficiently smooth such that all derivatives that we write exist and are continuous.

% We assume that the functions are smooth enough so the differentiations we write exist.

% Let $\rho>0$, and consider the function $\mbox{softmin}_\rho : \BBR \times \cdots \times \BBR \to \BBR$ defined by 
% \begin{equation}\label{eq:softmin}
% \mbox{softmin}_\rho (z_1,\ldots,z_N) \triangleq -\frac{1}{\rho}\log\sum_{i=1}^Ne^{-\rho z_i},
% \end{equation}
% which is the \textit{soft minimum}. The next result relates the soft minimum to the minimum.
% \begin{fact} \label{fact:softmin_bound}
% \rm{
% Let $z_1,\ldots, z_N \in \BBR$. 
% Then,
% \begin{align*}
%  \min \, \{ z_1,\ldots, z_N \} - \frac{\log N }{\rho} 
%  &\le \mbox{softmin}_\rho(z_1,\ldots,z_N) \\
%  &< \min \, \{z_1,\ldots, z_N\}.
% \end{align*}
% }
% \end{fact}

\begin{figure}[t]
    \centering
    \includegraphics[width=0.35\textwidth]{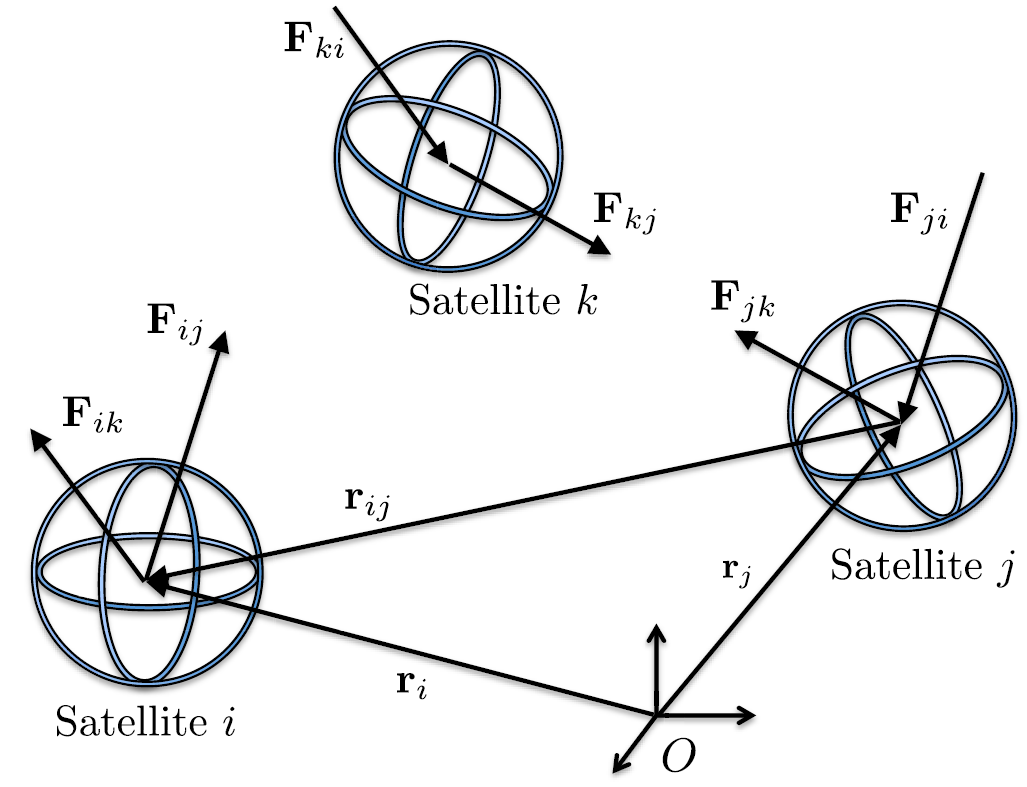}
    \caption{Each satellite is equipped with an electromagnetic actuation system consisting of three orthogonal coils. }
    \vspace{-10pt}
    \label{fig:Satellites_formation}
\end{figure}

\section{Problem Formulation}
\label{sec:Problem_Formulation}

Consider a system of $n$ satellites, where each satellite has mass $m$, as shown in Fig.~\ref{fig:Satellites_formation}.
The position $\mathbf r_i$ locates the mass center of satellite $i$ relative to the origin of an inertial frame $\CMcal{F}$ that consists of the right-handed set of mutually orthogonal unit vectors $\begin{bmatrix} \mathbf{i} &\mathbf{j} &\mathbf{k} \end{bmatrix}$. 
The velocity $\mathbf{v_i}$ and acceleration $\dot{\mathbf{v}}_i$ are the first and the second time-derivatives of $\mathbf{r}_i$ with respect to $\CMcal{F}$. 
The relative position $\mathbf{r}_{ij} \triangleq \mathbf{r}_i - \mathbf{r}_j$ locates the mass center of satellite $i$ with respect to the mass center of satellite $j$. 
The relative velocity $\mathbf{v}_{ij}$ is the time derivative of $\mathbf{r}_{ij}$ with respect to $\CMcal{F}$.

Each satellite is equipped with an electromagnetic actuation system that consists of 3 orthogonal electromagnetic coils, which are driven by a current source to generate a controllable electromagnetic field. 
These electromagnetic fields interact to produce intersatellite forces, which can control the satellites' relative positions. 
More information on electromagnetic actuation for satellites is available in \cite{Kong2004,KWON2010,Porter2014,Abbasi2022}.
Each coil creates a magnetic dipole, and the resulting intersatellite force applied to satellite $i$ by satellite $j$ is given by
\begin{align}
  \mathbf{F}_{ij} \triangleq \frac{3 \mu_0}{4 \pi |\mathbf r_{ij}|^4} \mathbf{f} (\mathbf{r}_{ij}, \mathbf{u}_i, \mathbf{u}_j), \label{F_ij}
\end{align}
where $\mu_0$ is the vacuum permeability constant, $| \mathbf{r}_{ij} |$ is the magnitude of  $\mathbf{r}_{ij}$, $\mathbf{u}_i$ is the magnetic moment of satellite $i$, and 
\begin{align}
    \mathbf{f}(\mathbf{r}_{ij},&\mathbf{u}_{i},\mathbf{u}_j) \triangleq  
    \left( \mathbf{u}_j \cdot \frac{\mathbf{r}_{ij}}{|\mathbf{r}_{ij}|} \right) \mathbf{u}_i + 
 \left(\mathbf{u}_i \cdot \frac{\mathbf{r}_{ij}}{|\mathbf{r}_{ij}|} \right) \mathbf{u}_j 
 \notag 
   \\
 &+ \left[(\mathbf{u}_i \cdot \mathbf{u}_j)-5 \left(\mathbf{u}_i \cdot \frac{\mathbf{r}_{ij}}{|\mathbf{r}_{ij}|} \right) \left(\mathbf{u}_j \cdot \frac{\mathbf{r}_{ij}}{|\mathbf{r}_{ij}|} \right) \right]\frac{\mathbf{r}_{ij}}{|\mathbf{r}_{ij}|} \label{f(r,ui,uj)}.
\end{align}
See \cite{Ahsun2006,Schweighart2010} for details on the force model (\ref{F_ij}) and (\ref{f(r,ui,uj)}). 
Since $\mathbf{r}_{ij} = -\mathbf{r}_{ji}$, it follows from (\ref{F_ij}) and (\ref{f(r,ui,uj)}) that $\mathbf{F}_{ij} = - \mathbf{F}_{ji}$, that is, the force applied to satellite $j$ by satellite $i$ is equal and opposite the force applied to $i$ by $j$. 
The magnetic moment $\mathbf{u}_i$ is a function of the current supplied to the electromagnetic coils, and this is the control for satellite $i$.

Thus, the translational dynamics of satellite $i$ are
\begin{align}
    \dot{\mathbf{v}}_i = \frac{c_0}{m} \sum_{j \in \CMcal{I} \setminus \{ i \}}\frac{1}{|\mathbf{r}_{ij}|^4}\mathbf{f}(\mathbf{r}_{ij},\mathbf{u}_{i},\mathbf{u}_j)
    \label{accel_i}
\end{align}
where $c_0 = 3 \mu_0 / (4 \pi)$. 
Since $\mathbf{F}_{ij} = - \mathbf{F}_{ji}$, it follows from (\ref{accel_i}) that $ \sum_{i \in \CMcal{I}} \dot{\mathbf{v}}_i (t) = 0$, which implies that the linear momentum of the system consisting of all satellites is conserved. 
Thus, the intersatellite electromagnetic forces can be used to alter relative positions, but they have no effect on the overall mass center of the satellites.

%%%%%%%%%%%%%%%%%%%%%%%%%%%%%%%%%%%%%%%%%%%%%%%%%%%%%%%%%%%%%%%%%%%%%%%%%%%%%%%%%%%%%%%%%%%%%%%%%%%%%%%%%%%%%%%%%%%%%%%%%%%%%%%%%%%%%%%%%%%%%%%%%%%%%%%%%%%%%%%%%%%%%%%%%%%%%%%%%%%%%%%%%%%%%%%%%%%%%%%%

\subsection{Piecewise-Sinusoidal Controls}
\label{sub_sec:Piecewise-Sinusoidal Controls}

This section reviews the piecewise-sinusoidal magnetic-moment approach in~\cite{Abbasi2022}, which is used to address the complex coupling that occurs between the electromagnetic fields generated by all satellites. 
In this approach, each satellite uses a piecewise-sinusoidal magnetic moment $\mathbf{u}_i$, consisting of a sum of $n-1$ piecewise-continuous sinusoids with $n-1$ unique frequencies, where each unique frequency is common to only one pair of satellites. 
Thus, there are a total of $n(n-1)/2$ unique frequencies.
Using this approach, the amplitudes of each sinusoid pair that share a common unique frequency can be selected to prescribe the average intersatellite force between the associated satellite pair.

For each satellite pair $(i, j ) \in \mathcal{P}$, we assign an interaction frequency $\omega_{ij} > 0$, where $\omega_{ij} = \omega_{ji}$ is unique. 
Let $T > 0$ be a common multiple of $\{ 2\pi/\omega_{ij}: (i,j) \in \SP \}$. 
Next, for all $k \in \mathbb{N}$ and $t \in [ kT, kT + T )$, we consider the piecewise-sinusoidal control
\begin{align}
    \mathbf{u}_i(t)=\sum_{j \in \mathcal{I} \setminus \{i\} } \mathbf p_{ij,k} \sin \omega_{ij}t,
    \label{eqn:u_i}
\end{align}
where the amplitude sequences $\{ \mathbf{p}_{ij,k} \}_{k=0}^\infty$ are now the control variables. 
Notice that satellite $i$ now has $n-1$ control sequences, namely, $\{ \mathbf{p}_{ij,k} \}_{k=0}^\infty$ for all $j \ne i$.

The following result demonstrates that the piecewise-sinusoidal control \eqref{eqn:u_i} approximately decouples the time-averaged force between each pair of satellites. 
This result is in~\cite[Prop. 1]{Abbasi2022}.

\begin{proposition}\label{Proposition_1}
Consider $\mathbf{f}$ and $\mathbf{u}_i$ given by (\ref{f(r,ui,uj)}) and (\ref{eqn:u_i}). 
Let $\mathbf{r}$ be a constant vector, and let $i, j \in \mathbb{N}$.
Then, for all $k \in \mathbb{N}$, 
\begin{equation}
        \frac{1}{T} \int^{kT+T}_{kT} \mathbf f(\mathbf{r}, \mathbf{u}_i(t),\mathbf{u}_j(t) ) \, {\rm d}t = \frac{1}{2} \mathbf f(\mathbf{r}, \mathbf{p}_{ij,k},\mathbf{p}_{ji,k}).
    \notag
\end{equation}
\end{proposition}

\subsection{Time-Averaged Dynamics}

We use \Cref{Proposition_1} to develop an approximate model for \eqref{accel_i} and \eqref{eqn:u_i}. 
First, integrating \eqref{accel_i} over the interval $[kT, kT + T ]$ yields
\begin{equation}
    \mathbf{v}_i(kT+T) =  \mathbf{v}_i(kT) + \frac{T}{m} \sum_{j \in \mathcal{I} \backslash \{ i \}} \mathbf{\bar{F}}_{ij} (k),
\label{discrete_velocity}    
\end{equation}
where
\begin{equation}
    \bar{\mathbf{F}}_{ij} (k) \triangleq \frac{1}{T} \int_{kT}^{kT+T} \frac{c_0}{| \mathbf{r}_{ij} (t) |^4} \mathbf{f}(  \mathbf{r}_{ij} (t), \mathbf{u}_{i} (t), \mathbf{u}_{j} (t)) \, {\rm d}t,
    \label{avg_F_ij}
\end{equation}
which is the average intersatellite force over $[kT, (k+1)T]$. 
For all $k \in \mathbb{N}$ and $t \in [ kT, kT + T )$, define the \textit{approximate average intersatellite force}
\begin{equation}
     \tilde{\mathbf{F}}_{ij} (t) \triangleq \frac{c_0}{2 | \mathbf{r}_{ij} (t) |^4} \mathbf{f}(  \mathbf{r}_{ij} (t), \mathbf{p}_{ij,k},\mathbf{p}_{ji,k}).
     \label{avg_approx_F_ij}
\end{equation}
\Cref{Proposition_1} implies that if $\mathbf{r}_{ij} (t)$ is constant on $[ kT, kT + T )$, then $\tilde{\mathbf{F}}_{ij} (t) = \bar{\mathbf{F}}_{ij} (k)$ on $[ kT, kT + T )$. 
Thus, if $\mathbf{r}_{ij}$ does not change significantly over each period $T$, then \eqref{accel_i} and \eqref{eqn:u_i} can be approximated by 
\begin{equation}
    \dot{\tilde{\mathbf{v}}}_i = \frac{c_0}{2 m} \sum_{j \in \CMcal{I} \setminus \{ i \}}\frac{1}{|\tilde{\mathbf{r}}_{ij} \|^4}\mathbf{f}(\tilde{\mathbf{r}}_{ij},\mathbf{p}_{ij,k},\mathbf{p}_{ji,k}), \label{eq:avg_model}
\end{equation}
where $\tilde{\cdot}$ is the approximate variable.

Next, we resolve \eqref{eq:avg_model} in the inertial frame $\CMcal{F}= \begin{bmatrix}
    \mathbf{i} &\mathbf{j} &\mathbf{k}
\end{bmatrix}$. 
Specifically, let $r_i \triangleq [\tilde{\mathbf{r}}_i]_{\CMcal{F}}$ and $v_i \triangleq [\tilde{\mathbf{v}}_i]_{\CMcal{F}}$, and it follows from \eqref{eq:avg_model} that 
\begin{align}
    \dot{r}_{i} &= v_{i}, \label{eq:model.1}\\
    \dot{ v }_{i} &= \frac{c_0}{2m} \sum_{j \in \CMcal{I} \setminus \{ i \}}\frac{1}{ \norm{r_{ij}}^4 } f( r_{ij}, p_{ij}, p_{ji}), \label{eq:model.2}
\end{align}
where $r_{ij} \triangleq r_{i} - r_{j}$, and
\begin{align}
    f( r_{ij}, p_{ij}, p_{ji}) &\triangleq \frac{ p^{\mathrm{T}}_{ji} r_{ij}}{ \norm{r_{ij}} } p_{ij} + \frac{p^{\mathrm{T}}_{ij}  r_{ij}}{\norm{r_{ij}}} p_{ji} +  \frac{ p^{\mathrm{T}}_{ij} p_{ji}}{ \norm{r_{ij}} } r_{ij}
 \notag 
   \\
 &\qquad 
 - 5 \frac{ p^{\mathrm{T}}_{ij} r_{ij} p^{\mathrm{T}}_{ji} r_{ij} }{ \norm{r_{ij}}^3 } r_{ij},
\label{eq:model.3}
\end{align}
and $p_{ij} \colon [0,\infty) \to \BBR^3$ are the controls, which are sampled in order to generate the amplitudes in \eqref{eqn:u_i}.
Specifically, the sinusoidal amplitudes are 
\begin{equation}
    \mathbf p_{ij,k} = \begin{bmatrix}
    \mathbf{i} &\mathbf{j} &\mathbf{k}
\end{bmatrix} p_{ij}(kT). \label{eq:p_ij_k}
\end{equation}
Note that~\cite[Sec. V]{Abbasi2022} shows that the amplitude pair $(p_{ij},p_{ji})$ can be selected such that $f( r_{ij}, p_{ij}, p_{ji})$ takes on any prescribed value. 
Thus, this paper uses \cref{eq:model.1,eq:model.2} to design a desired $f( r_{ij}, p_{ij}, p_{ji})$ that achieves the formation-flying, state-constraint, and input-constraint objectives in the next subsection. 
Then, we compute an amplitude pair $(p_{ij},p_{ji})$ such that $f( r_{ij}, p_{ij}, p_{ji})$ is equal to its desired value.

\subsection{Control Objective}

Let $d_{ij} \in \BBR^3$ be the desired position of satellite $i$ relative to satellite $j$ resolved in the inertial frame $\CMcal{F}$.
Note that $d_{ji} = -d_{ij}$.
The formation control objective is to design piecewise-sinusoidal magnetic-moment control \eqref{eqn:u_i} such that the relative positions ${r}_{ij}$ converge to the desired relative positions ${d}_{ij}$, that is, the satellites move to the desired formation.

In addition to the formation objective, the satellites must satisfy state and control-input constraints.  
First, let $\ubar{r} > 0$ denote the collision radius, that is, the minimum acceptable distance between any 2 satellites. 
Next, let $\bar{v} > 0$ denote the maximum intersatellite speed, that is, the maximum acceptable relative speed between any 2 satellites. 
Finally, let $\bar Q > 0$ denote the maximum apparent power capability of each satellite. 
We note that the apparent power of the $i$th satellite is approximately $\frac{1}{N^2 A^2} \sum_{j \in \SI\backslash \{ i \} } {Z_{ij}} \| p_{ij} \|^2$, where $Z_{ij} >0$ is the impedance of the coils at frequency $\omega_{ij}$, $A>0$ is the coil cross-sectional area, and $N>0$ is the number of turns in the coil. 
In summary, the control objectives are:
\begin{enumerate}[leftmargin=0.8cm]
	\renewcommand{\labelenumi}{(O\arabic{enumi})}
	\renewcommand{\theenumi}{(O\arabic{enumi})}

\item\label{obj1}
For all $(i,j) \in \SP$, $\lim_{t \to \infty} {r}_{ij}(t) = {d}_{ij}$.

\item\label{obj2}
For all $(i,j) \in \SP$ and all $t \ge 0$, $\| {r}_{ij}(t) \| \ge \ubar{r}$.

\item\label{obj3}
For all $(i,j) \in \SP$ and all $t \ge 0$, $\| {v}_{i}(t) - v_j(t) \| \le \bar{v}$.

\item\label{obj4}
For all $i \in \SI$ and all $t \ge 0$, $$ \frac{1}{N^2 A^2} \sum_{j \in \SI\backslash \{ i \} } Z_{ij} \| p_{ij}(t) \|^2 \le \bar Q.$$

\end{enumerate}

\section{Control Amplitude Pair to Achieve Prescribed Intersatellite Force} \label{Section:Allocation}

This section provides a construction for the amplitude pair $(p_{ij},p_{ji})$ such that $f( r_{ij}, p_{ij}, p_{ji})$ takes on a prescribed value. 
The construction in this section differs from the one provided in~\cite{Abbasi2022}, because the construction in~\cite{Abbasi2022} cannot be applied with the input constraint \ref{obj4}.

First, consider $a_x, a_y, b_x, b_y \colon \mathbb{R}^3 \times \mathbb{R}^3 \rightarrow \mathbb{R}^3$ given by  
\begin{align}
    a_x(r,f_*) &\triangleq -\frac{\sgn(r^{\mathrm{T}} f_*)}{2} 
     \left( \frac{|r^{\mathrm{T}} f_*| + \Phi_1(r,f_*) }{\norm{r}} \right)^{\frac{1}{2}}, \label{eq:ax}
    \\
    a_y(r,f_*) &\triangleq \frac{\sgn \left( \Phi_2(r,f_*) \right)}{ \sqrt{2} }  \left( \dfrac{ -|r^{\mathrm{T}} f_*| + \Phi_3(r,f_*) }{\norm{r}} \right)^{\frac{1}{2}}, \label{eq:ay}
    \\
    b_x(r,f_*) &\triangleq  \frac{1}{2} \left( \dfrac{ |r^{\mathrm{T}} f_*| +\Phi_3(r,f_*) }{  \norm{r} } \right)^{ \frac{1}{2} }, \label{eq:bx}
    \
    \\
    b_y(r,f_*) &\triangleq - \frac{\sgn \left( (r^{\mathrm{T}} f_*) \Phi_2(r,f_*) \right) }{ \sqrt{2} }  \bigg( \dfrac{ -|r^{\mathrm{T}} f_*|}{\norm{r}} 
    \nn
    \\
     &\qquad + \dfrac{ \Phi_1(r,f_*) }{\norm{r}} \bigg)^{ \frac{1}{2} }, \label{eq:by}
\end{align}
where $\Phi_1, \Phi_2, \Phi_3 \colon \mathbb{R}^3 \times \mathbb{R}^3 \rightarrow \mathbb{R}^3$ are define by
\begin{align}    
 \Phi_1(r,f_*) &\triangleq \sqrt{2 \norm{r}^2 \norm{f_*}^2 - (r^{\mathrm{T}}  f_*)^2 },\label{eq:Phi1}
 \\
 \Phi_2(r,f_*) &\triangleq \sqrt{\norm{r}^2 \norm{f_*}^2 - (r^{\mathrm{T}}  f_*)^2 },\label{eq:Phi2}
 \\
 \Phi_3(r,f_*) &\triangleq \sqrt{2 (2-\sgn(r^{\mathrm{T}} f_*)^2 )^2 \norm{r}^2 \norm{f_*}^2 - (r^{\mathrm{T} } f_*)^2 }.\label{eq:Phi3}
\end{align}
Next, consider $R: \mathbb{R}^3 \times \mathbb{R}^3 \rightarrow \mathrm{SO}(3)$ given by
\begin{align}
    R(r, f_*) &\triangleq \begin{bmatrix}
             \dfrac{r^{\mathrm{T}}}{\norm{r}}
        \\
          \dfrac{  r^{\mathrm{T}} r f_*^{\mathrm{T}} - r^{\mathrm{T}} f_* r^{\mathrm{T}}  }{ \norm{r} \Phi_2(r,f_*) }
        \\
         \dfrac{ \left( [r]_{\times}  f_* \right)^{\mathrm{T}} }{  \Phi_2(r,f_*) } 
    \end{bmatrix}
    \label{eq:R}
\end{align}
which is the rotation from $\mathcal{F}$ to a frame whose first axis is parallel to $r_{ij}$, and whose remaining axes are orthogonal to $r_{ij}$ and $f_{ij}$.
Finally, consider $c_1,c_2 \colon \mathbb{R}^3 \times \mathbb{R}^3 \rightarrow \mathbb{R}$ given by
\begin{align}
    c_1(r,f_*) &\triangleq R(r,f_*)^{\mathrm{T}} \begin{bmatrix}
        a_x(r,f_*)\\
        a_y(r,f_*)\\
        0
    \end{bmatrix}, \label{eq:c1}
    \\
    c_2(r,f_*) &\triangleq R(r,f_*)^{\mathrm{T}} \begin{bmatrix}
        b_x(r,f_*)\\
        b_y(r,f_*)\\
        0
    \end{bmatrix}. \label{eq:c2}
\end{align}
The next result shows that $(c_1,c_2)$ is a construction for the amplitude pair $(p_{ij},p_{ji})$ such that $f( r_{ij}, p_{ij}, p_{ji})$ takes on a prescribed value $f_*$. 
The proof follows from using \Cref{eq:ax,eq:ay,eq:bx,eq:by,eq:Phi1,eq:Phi2,eq:Phi3,eq:R,eq:c1,eq:c2} to evaluate $f(r, c_1(r,f_*), c_2(r,f_*) )$ and is omitted for brevity.

\begin{proposition}    \label{Proposition_2}
    For all $r \in \mathbb{R}^3  \backslash \{ 0 \}$ and $f_* \in \mathbb{R}^3$, $f(r, c_1(r,f_*), c_2(r,f_*)) = f_*$.
\end{proposition}

Next, let $f_{ij} \colon [0,\infty) \to \BBR^3$ be the prescribed value for the intersatellite-force function $f(r_{ij}, p_{ij}, p_{ji})$. 
Then, let 
 \begin{equation}
     p_{ij} = \begin{cases}
            c_1(r_{ij}, f_{ij}), &i<j,\\
            c_2(r_{ij}, f_{ij}), &i>j,\\
        \end{cases}
        \label{eqn:p_ij_allocation}
 \end{equation}
 and it follows from \Cref{Proposition_2} that $f(r_{ij}, p_{ij}, p_{ji}) = f_{ij}$. 
 In other words, the amplitude pair $(p_{ij},p_{ji})$ given by \eqref{eqn:p_ij_allocation} achieves the prescribed intersatellite force $f_{ij}$. 
 The next section presents a optimal closed-form feedback control for $f_{ij}$ that satisfies \ref{obj1}--\ref{obj4}.

The next result relates the magnitudes of $c_1(r,f_*)$ and $c_2(r,f_*)$.
This result is useful for addressing \ref{obj4}. 
The proof following from direct computation and is omitted for brevity. 

\begin{proposition}    \label{Proposition_mag_allo_mat}
Let $r \in \mathbb{R}^{3} \backslash \{0 \}$, and $f_* \in \mathbb{R}^3$. Then the following statements hold:
    \begin{enumerate}
        \item If $r^{\mathrm{T}} f_* \neq 0$, then $\norm{c_1(r,f_*)} = \norm{c_2(r,f_*)}$. \label{Prop_4:item_1}
        \item If $r^{\mathrm{T}} f_* = 0$, then  $\norm{c_1(r,f_*)}^2 = \frac{1}{\sqrt{2}} \norm{ c_2(r, f_*) }^2$. \label{Prop_4:item_2}
    \end{enumerate}
\end{proposition}

\section{EMFF with State and Input Constraints}
\label{sec:Formation control with state and input constraints}

The state $x \colon [0,\infty) \to \BBR^{6n}$ and intersatellite-force function $\nu \colon [0,\infty) \to \BBR^{\ell_\nu}$ for the entire satellite system are given by
\begin{equation}
    x(t) \triangleq \begin{bmatrix}
        r_1(t)\\
        r_2(t)\\
        \vdots\\
        r_n(t) \\
        v_1(t)\\
        v_2(t)\\
        \vdots\\
        v_n(t) \\
    \end{bmatrix}, \qquad \nu(t) \triangleq \begin{bmatrix}
        f_{12}(t)\\
        \vdots\\
        f_{1n}(t)\\
        f_{23}(t)\\
        \vdots\\
        f_{2n}(t)\\
        \vdots\\
        f_{n-2, n-1}(t)\\
        f_{n-2, n}(t)\\
        f_{n-1, n}(t)
    \end{bmatrix}, 
\end{equation}
where $\ell_\nu \triangleq 3n(n-1)/2$.
Thus, \Cref{eq:model.1,eq:model.2,eq:model.3,eqn:p_ij_allocation} and \Cref{Proposition_2} imply that 
\begin{equation}
    \dot{x}(t) = A x(t) + B \zeta(x(t),\nu(t)),
    \label{eqn:State_space_n_satellites}
\end{equation}
where $x(0) = x_0 \in \BBR^{6n}$,
\begin{gather}
  A \triangleq \begin{bmatrix}
        0_{n \times n} &I_n\\
        0_{n \times n} &0_{n \times n}
    \end{bmatrix} \otimes I_3, 
    \quad
B \triangleq \frac{c_0}{2m} \begin{bmatrix}
        0_{3n \times \ell_v}\\ B_0 \otimes I_{3} 
    \end{bmatrix},
    \label{eq:A}\\
    % \quad B \triangleq \begin{bmatrix}
    %     0\\
    %     I
    % \end{bmatrix}, \quad g(x) \triangleq    
    %    \left ( M \gamma(x) \right )  \otimes I_{3} ,\\
%JBH: Please insert another column in B before the ellipse 
% SSK: I don't know what you mean.
% JBH: The complicated matrix in B has 2 block colums followed by ellise (...) and a last column. To help the reader understand the pattern, you should add anothe block column before ...
%SSK: Added a block column before the last column
%
% JBH: I split to clean up. Please check
% SSK: Checked.
    B_0 \triangleq \begin{bsmallmatrix}
        {1}^{\mathrm{T}}_{n-1} &0_{1 \times (n-2)}  &\dots &0_{1 \times 2} &0
        \\
        -I_{n-1} &\begin{bsmallmatrix} {1}^{\mathrm{T}}_{n-2} \\
        -I_{n-2} \end{bsmallmatrix} &\dots &\begin{bsmallmatrix} 
        0_{(n-4) \times 2} \\
        1^{\mathrm{T}}_{2} \\
        -I_2 
        \end{bsmallmatrix}
        &\begin{bsmallmatrix} 
        0_{n-3} \\
        1 \\
        -1 
        \end{bsmallmatrix} \end{bsmallmatrix} , \label{eq:B}
\end{gather}
and $\zeta \colon \BBR^{6n} \to \BBR^{\ell_\nu}$ is defined as 
\begin{equation}
\zeta(x,\nu) \triangleq \left ( \gamma(x) \otimes I_3 \right ) \nu, \label{eq:mu}\\
\end{equation}
where $\gamma \colon \BBR^{6n} \to \BBR^{n(n-1)/2 \times n(n-1)/2}$ is defined as
\begin{align}
\gamma(x)  &\triangleq \mathrm{diag} \Big ( \norm{r_{12}}^{-4}, \hdots,\norm{r_{1n}}^{-4},\norm{r_{23}}^{-4},\hdots,\norm{r_{2n}}^{-4},\nn\\
    & \hdots ,\norm{r_{n-2,n-1}}^{-4},\norm{r_{n-2,n}}^{-4},\norm{r_{n-1,n}}^{-4} \Big ), \label{eq:gamma}
\end{align} 
and ${1}_{n} \triangleq \begin{bmatrix}
    1 &1 &\dots &1 
\end{bmatrix}^{\mathrm{T}} \in \mathbb{R}^n$, $\otimes$ denotes Kronecker product, and $\mbox{diag}(\cdot)$ denotes a diagonal matrix whose diagonal elements are the arguments. 

\begin{figure*}[t]
  \includegraphics[width=1\textwidth,clip=true,trim= 0.1in 0.1in 0.1in 0.1in]{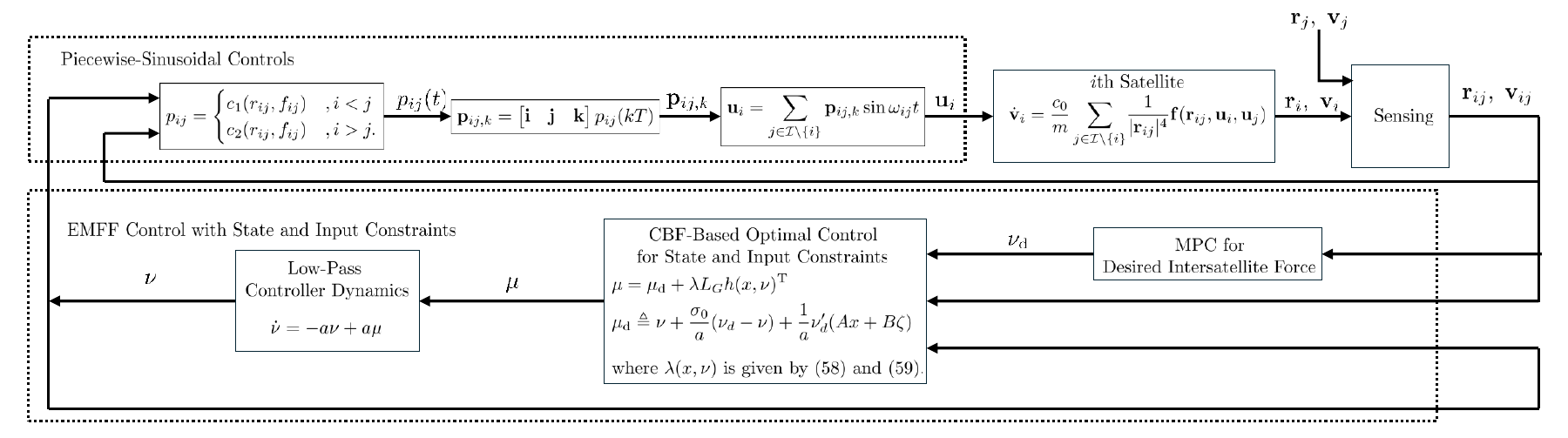}
  \centering
  \caption{EMFF control with state and input constraints using AMFF.}
  \label{fig:block_diagram_cbf_emff}
\end{figure*}

\Cref{fig:block_diagram_cbf_emff} illustrates the EMFF control architecture with state and input constraints. 
The approach consists of an MPC module for computing desired intersatellite forces; a CBF-Based Optimization with controller dynamics for satisfying state and input constraints; and the piecewise-sinusoidal control construction given in \Cref{Section:Allocation}. 
The following subsections present the MPC module and the CBF-based optimization.

\vspace{-0.5em}
% We note that \cref{eqn:State_space_n_satellites} is linear in $x$ and $\zeta$. 

\subsection{Desired Formation Control}
\label{sec:MPC}
Let $t_\rmf > 0$, and let $W_{ij},W_i \in \BBR^{3 \times 3}$ and $W_\zeta \in \BBR^{\ell_\nu \times \ell_\nu}$ be positive definite.
Then, for all $x_\rmi \in \BBR^{6n}$, consider the cost function 
\begin{align}
    J(x_{\rmi},\hat \zeta ( \cdot ) ) &\triangleq \int_{0}^{T_{\mathrm{f}}} \mspace{-10mu} \sum_{(i,j) \in \SP} \left( r_{ij}(\tau) - d_{ij} \right)^\rmT W_{ij} \left ( r_{ij}(\tau) - d_{ij} \right) \nn \\
    &\qquad + \sum_{i \in \SI} (v_i(\tau)-v_j(\tau))^\rmT W_i (v_i(\tau) -v_j(\tau)) \nn \\ 
    &\qquad+ \hat \zeta(\tau)^\rmT W_\zeta \hat \zeta(\tau) \, \mathrm{d} \tau \label{eqn:MPC_cost_fn}
\end{align}
subject to the dynamics \eqref{eqn:State_space_n_satellites} with $x(0) = x_\rmi$ and $\zeta = \hat \zeta$. 
For each $x_\rmi \in \BBR^{3n}$, let $\zeta_*(x_\rmi,\cdot) \colon [0,T_\rmf) \to \BBR^{\ell_\nu}$ denote the minimizer of \eqref{eqn:MPC_cost_fn} subject to $\dot x = A x + B \hat \zeta$ with initial condition $x_\rmi$. 
Then, we consider the desired feedback law $\zeta_\rmd \colon \BBR^{6n} \to \BBR^{\ell_\nu}$, defined by $\zeta_\rmd(x) \triangleq \zeta_*(x,0)$.
The control $\zeta_{\rmd}(x)$ is computed numerically in real time using standard linear MPC (e.g., \cite{Iskandarani2014}). 
%We note that if $T_\rmf = \infty$, then $\zeta_{\rmd}(x)$ is the linear-quadratic regulator. 
% = K_{\rm lqr} x$, where $K_{\mathrm{lqr}} \in \mathbb{R}^{\ell_\nu \times 6n}$ is the linear-quadratic-regulator gain.
Next, we define the \textit{desired intersatellite-force function}
\begin{equation}
    \nu_\rmd(x) \triangleq \left ( \gamma(x) \otimes I_3 \right )^{-1} \zeta_\rmd(x), \label{eq:nu_d}
\end{equation}
and it follows from \eqref{eq:mu} that $\nu = \nu_\rmd$ yields $\zeta = \zeta_\rmd$. 
The desired intersatellite-force function $\nu_\rmd$ is designed to achieve \ref{obj1}; however, this feedback control does not address \ref{obj2}--\ref{obj4}. 
The remainder of this section presents a method for constructing the control $\nu$ that is as close as possible to $\nu_\rmd$, while satisfying nonlinear state constraints (i.e., \ref{obj2} and \ref{obj3}) and nonlinear input constraints (i.e., \ref{obj4}).

\subsection{Barrier Functions for Constraints}

To address the state constraint \ref{obj2}, consider the continuously differentiable candidate CBF
\begin{equation}
    R_{ij}(x) \triangleq \frac{1}{2} \left ( \|r_{ij}\|^2 - \ubar{r}^2 \right ). \label{eq:Rij}
\end{equation}
The intersection of these functions' zero-superlevel sets is
\begin{equation}
    \SR \triangleq \{ x \in \BBR^{6n} \colon \mbox{for all } (i,j) \in \SP, R_{ij}(x) \ge 0 \}.
\end{equation}
Note that $R_{ij}(x) \ge 0$ if and only if satellite $i$ and $j$ are not inside the collision radius (i.e., $\|r_{ij}\|\ge \ubar{r}$). 
Thus, \ref{obj2} is satisfied if and only if for all $t \ge 0$, $x(t) \in \SR$.
The candidate CBF $R_{ij}$ has relative degree $2$ with respect to the \Cref{eqn:State_space_n_satellites,eq:A,eq:B,eq:mu,eq:gamma} where the input is $\nu$. 

To address the state constraint \ref{obj3}, consider the continuously differentiable candidate CBF
\begin{equation}
    V_{ij}(x) \triangleq \frac{1}{2} \left ( \bar{v}^2 - \|v_{ij}\|^2 \right ). \label{eq:Vij}
\end{equation}
The intersection of these functions' zero-superlevel sets is
\begin{equation}
    \SV \triangleq \{ x \in \BBR^{6n} \colon \mbox{for all } (i,j) \in \SP, V_{ij}(x) \ge 0 \}.
\end{equation}
Thus, \ref{obj3} is satisfied if and only if for all $t \ge 0$, $x(t) \in \SV$.
The candidate CBF $V_{ij}$ has relative degree one with respect to the \Cref{eqn:State_space_n_satellites,eq:A,eq:B,eq:mu,eq:gamma} where the input is $\nu$.

Objective \ref{obj4} is an input constraint related to the magnitude of the amplitude $p_{ij}$. 
However, \Cref{eqn:p_ij_allocation} and \Cref{Proposition_mag_allo_mat} imply that $\| p_{ij} \|^2$ is not continuous in $(r_{ij},f_{ij})$.
Thus, we consider a continuously differentiable function $\psi: \mathbb{R}^{3} \times \mathbb{R}^{3} \rightarrow \mathbb{R}$ that upper bounds $\| p_{ij} \|^2$. 
Specifically, define
\begin{align}
    \psi(r_{ij}, &f_{ij}) \triangleq -\frac{1}{4} \left(\frac{ r_{ij}^{\mathrm{T}} f_{ij} }{\norm{r_{ij}} } \right) \tanh \left( \frac{r_{ij}^{\mathrm{T}} f_{ij}}{\epsilon_1 \norm{r_{ij}}} \right) 
    \notag
    \\
    &+ \frac{\sqrt{ 2 \norm{ r_{ij} }^2 \norm{f_{ij}}^2 - \left( r_{ij}^{\mathrm{T}} f_{ij} \right)^2  + \epsilon_2 \norm{ r_{ij} }^2 } }{\norm{r_{ij}}},&&
    \label{eqn:psi(r,f_*)}
\end{align}
where $0<\epsilon_1<<1$ and $0<\epsilon_2<<1$.
The next result combined with \Cref{eqn:p_ij_allocation} shows that $\psi(r_{ij}, f_{ij}) \ge \| p_{ij} \|^2$.

\begin{proposition}
Let $r \in \mathbb{R}^{3} \backslash \{ 0 \}$ and $f_* \in \mathbb{R}^3$.
Then, $\psi(r, f_*) > \norm{c_1(r,f_*)}^2 \ge \norm{c_2(r,f_*)}^2$. 
    \label{Proposition_upper_bound_mag}
\end{proposition}

% \begin{proposition}
%     Let $r \in \mathbb{R}^{3} \backslash \{ 0 \}$ and $f_* \in \mathbb{R}^3$, then the following statements hold:
%     \begin{enumerate}
%         \item If $r^{\mathrm{T}} f_* \neq 0$, then $\psi(r, f_*) > \norm{c_1(r,f_*)}^2 = \norm{c_2(r,f_*)}^2$. \label{Prop_5:item_1}
%         \item If $r^{\mathrm{T}} f_* = 0$, then $\psi(r, f_*) > \norm{c_1(r,f_*)}^2 > \norm{c_2(r,f_*)}^2$. 
%         \label{Prop_5:item_2}
%     \end{enumerate}
%     \label{Proposition_upper_bound_mag}
% \end{proposition}

To address the state and input constraint \ref{obj4}, we consider the continuously differentiable function 
\begin{equation}
    Q_i(x,\nu) \triangleq \bar Q - \frac{1}{N^2 A^2}\sum_{j \in \SI\backslash \{ i \} } Z_{ij} \psi(r_{ij}, f_{ij}).\label{eq:Qi} 
\end{equation}
The intersection of these functions' zero-superlevel sets is
\begin{equation}
    \SQ \triangleq \{ (x,\nu) \in \BBR^{6n} \times \BBR^{\ell_\nu} \colon \mbox{for all } i \in \SI, Q_{i}(x) \ge 0 \}.
\end{equation}
\Cref{Proposition_upper_bound_mag} and \Cref{eqn:p_ij_allocation} imply that if for all $t \ge 0$, $(x(t),\nu(t)) \in \SQ$, then \ref{obj4} is satisfied. 
However, \eqref{eq:Qi} depends explicitly on the control $\nu$, which means that \eqref{eq:Qi} is not a valid candidate CBF.

Next, we define 
\begin{equation}
    \SSS_\rms \triangleq \{ (x,\nu) \in \BBR^{6n} \times \BBR^{\ell_\nu} \colon x \in \SR \cap \SV \mbox{ and } (x,\nu)  \in \SQ \},
\end{equation}
and it follows that if for all $t \ge 0$, $(x(t),\nu(t)) \in \SSS_\rms$, then \ref{obj2}--\ref{obj4} are satisfied. 
The following subsection introduces control dynamics in order to make $Q_i$ into a candidate CBF.
Then, we use a soft-minimum function to compose \Cref{eq:Rij,eq:Vij,eq:Qi} into one relaxed CBF that can be used as a constraint to ensure that for all $t \ge 0$, $(x(t),\nu(t)) \in \SSS_\rms$.

%\Cref{eq:ax,eq:ay,eq:bx,eq:by,eq:Phi1,eq:Phi2,eq:Phi3,eq:R,eq:c1,eq:c2,eqn:p_ij_allocation} 

\subsection{Control Dynamics and Soft-Minimum Relaxed CBF for State and Input Constraints}

We address the state and input constraints (i.e., \ref{obj2}--\ref{obj4}) by applying the method in \cite{rabiee2024b,rabiee2024a}.
The method in \cite{rabiee2024b,rabiee2024a} uses control dynamics to transform the input constraint into a controller-state constraints, and uses a soft-minimum function to compose multiple candidate CBFs (one for each state and input constraint) into a single relaxed CBF.
%; (3) a desired input to the control dynamics; and (4) a minimum-intervention quadratic program. 

First, let the control $\nu$ be generated by the first-order control dynamics
\begin{equation}
    \dot{\nu}(t) = -a  \nu(t) + a \mu(t), \label{eqn:controller_state_space} 
\end{equation}
where $a>0$, $\nu(0) =\nu_0 \in \BBR^{\ell_\nu}$ is the initial condition, and $\mu \colon [0\infty) \to \BBR^{\ell_\nu}$ is the surrogate control, that is, the control input to \eqref{eqn:controller_state_space}.

Next, consider the \textit{desired surrogate control} 
$\mu_\rmd \colon \BBR^{6n} \times \BBR^{\ell_\nu} \to \BBR^{\ell_\nu}$ defined by
\begin{equation}
    \mu_\rmd (x,\nu) \triangleq \nu + \frac{\sigma}{a}   [\nu_\rmd(x) - \nu] + \frac{1}{a} \nu_\rmd^\prime(x) [Ax + B \zeta(x,\nu) ], \label{eq:nu_d hat}
\end{equation}
where $\sigma > 0$.
As shown in \cite{rabiee2024b}, if $\mu= \mu_\rmd$, then $\nu$ converges exponentially to $\nu_\rmd$. 
In fact, \cite{rabiee2024b} implies that if $\mu= \mu_\rmd$ and $\nu_0 = \nu_\rmd(x_0)$, then $\nu(t) \equiv \nu_\rmd(x(t))$. 
In other word, $\mu_\rmd$ is the desired input to the control dynamics \eqref{eqn:controller_state_space} that yields the desired control $\nu_\rmd$ generated from the MPC in \Cref{sec:MPC}.

The cascade of \Cref{eqn:State_space_n_satellites,eq:A,eq:B,eq:mu,eq:gamma} and \Cref{eqn:controller_state_space} is 
\begin{equation}
\dot{\hat{x}} = \phi(\hat{x})  + G \mu, \label{eq:cascade_1}
\end{equation}
where 
\begin{equation}
    \hat{x} \triangleq \begin{bmatrix}
        x\\
        \nu
    \end{bmatrix}, \quad 
    \phi(\hat{x}) \triangleq \begin{bmatrix}
        A x + B\zeta(x,\nu)\\
        -a \nu
        \end{bmatrix}, \quad
    G \triangleq a \begin{bmatrix}
        0\\
        I
    \end{bmatrix}. \label{eq:cascade_2}
\end{equation}
Notice that the control dynamics \eqref{eqn:controller_state_space} make the control $\nu$ into a state of the cascade \Cref{eq:cascade_1,eq:cascade_2}. 
Thus, the constraint $Q_i(x(t),\nu(t)) > 0$, which is sufficient to satisfy \ref{obj4}, is transformed from an input constraint into a constraint on the state of the cascade. 
In fact, $Q_i$ has relative degree one with respect to \Cref{eq:cascade_1,eq:cascade_2} where the input is $\mu$. 
Also, note that $R_{ij}$ and $V_{ij}$ have relative degree 3 and 2 with respect to \Cref{eq:cascade_1,eq:cascade_2} where the input is $\mu$. 
Thus, the control dynamics \Cref{eqn:controller_state_space} increases the relative degree of the candidate CBFs $R_{ij}$ and $V_{ij}$ and converts $Q_i$ into a candidate CBF with relative degree one.

Since $R_{ij}$ and $V_{ij}$ have relative degree 3 and 2, we use a higher-order approach to construct higher-order candidate CBFs from $R_{ij}$ and $V_{ij}$.
First, let $\alpha_{0} \colon \mathbb{R} \to \mathbb{R}$ be a $3$-times continuously differentiable extended class-$\mathcal{K}$ function, and consider $R_{ij,1} \colon \mathbb{R}^n \to \mathbb{R}$ defined by
\begin{align}
R_{ij,1}(x) & \triangleq L_\phi R_{ij}(x) +\alpha_{0}(R_{ij}(x))\nn\\
&= R_{ij}^\prime(x) A x + \alpha_{0}(R_{ij}(x)). \label{eq:Rij_1}
\end{align}
Next, let $\alpha_{1} \colon \mathbb{R} \to \mathbb{R}$ be a $2$-times continuously differentiable extended class-$\mathcal{K}$ function, and consider $R_{ij,2} \colon \colon \mathbb{R}^n \times \BBR^{\ell_\nu} \to \mathbb{R}$ defined by
\begin{align}
R_{ij,2}(x,\nu) & \triangleq L_\phi R_{ij,1}(x) +\alpha_{1}(R_{ij,1}(x))\nn\\
&= \| v_{ij} \|^2 + R_{ij,1}^\prime(x) A B\zeta(x,\nu) + \alpha_{1}(R_{ij,1}(x)).
\label{eq:Rij_2}
\end{align}
Similarly, let $\alpha_{v} \colon \mathbb{R} \to \mathbb{R}$ be a $2$-times continuously differentiable extended class-$\mathcal{K}$ function, and consider $V_{ij,1} \colon \mathbb{R}^n \times \BBR^{\ell_\nu} \to \mathbb{R}$ defined by
\begin{align}
V_{ij,1}(x,\nu) & \triangleq L_\phi V_{ij}(x) +\alpha_{v}(V_{ij}(x))\nn\\
&= V_{ij}^\prime(x) B\zeta(x,\nu) + \alpha_{v}(V_{ij}(x)).
\label{eq:Vij_1}
\end{align}

% Specifically, let $\alpha_{v} \colon \mathbb{R} \to \mathbb{R}$ be a $2$-times continuously differentiable extended class-$\mathcal{K}$ function, and consider $V_{ij} \colon \mathbb{R}^n \times \BBR^{\ell_\nu} \to \mathbb{R}$ defined by
% \begin{align}
% V_{ij}(x,\nu) & \triangleq L_\phi V_{ij,0}(x) +\alpha_{v,0}(V_{ij,0}(x))\nn\\
% &= V_{ij,0}^\prime(x) B\zeta(x,\nu) + \alpha_{v,0}(V_{ij,0}(x)).
% \end{align}
% Similarly, let $\alpha_{0} \colon \mathbb{R} \to \mathbb{R}$ be a $3$-times continuously differentiable extended class-$\mathcal{K}$ function, let $\alpha_{r} \colon \mathbb{R} \to \mathbb{R}$ be a $3$-times continuously differentiable extended class-$\mathcal{K}$ function, and consider $R_{ij,1},R_{ij} \colon \mathbb{R}^n \times \BBR^{\ell_\nu} \to \mathbb{R}$ defined by
% \begin{align}
% R_{ij,1}(x) & \triangleq L_\phi R_{ij,0}(x) +\alpha_{0}(R_{ij,0}(x))\nn\\
% &= R_{ij,0}^\prime(x) A x + \alpha_{0}(R_{ij,0}(x)).
% \end{align}
% and 
% \begin{align}
% R_{ij}(x,\nu) & \triangleq L_\phi R_{ij,1}(x) +\alpha_{r}(R_{ij,1}(x))\nn\\
% &= 2 \| v_{ij} \|^2 + R_{ij,0}^\prime(x) A B\zeta(x,\nu) + \alpha_{r}(R_{ij,1}(x)).
% \end{align}
% Next, define the superlevel sets
% \begin{align}
% \SV &\triangleq \{ (x,\nu) \in \BBR^{6n} \times \BBR^{\ell_\nu} \colon \mbox{for all } (i,j) \in \SP, \nn\\
% &\qquad V_{ij,1}(x,\nu) \ge 0 \},\\
% %
% \SR_1 &\triangleq \{ x \in \BBR^{6n} \colon \mbox{for all } (i,j) \in \SP, R_{ij,1}(x) \ge 0 \},\\
% %
% \SR &\triangleq \{ (x,\nu) \in \BBR^{6n} \times \BBR^{\ell_\nu} \colon \mbox{for all } (i,j) \in \SP, \nn\\
% &\qquad R_{ij,2}(x,\nu) \ge 0 \}.
% \end{align}

Next, define 
\begin{equation}
    \SR_1 \triangleq \{ x \in \BBR^{6n} \colon \mbox{for all } (i,j) \in \SP, R_{ij,1}(x) \ge 0 \},
\end{equation}
which is the intersection of the zero-superlevel set of $R_{ij,1}$ for all $(i,j) \in \SP$.
Also, define 
\begin{align}
    \bar \SH \triangleq \{ (x,\nu) \in \SQ &\colon \mbox{for all } (i,j) \in \SP, R_{ij,2}(x,\nu) \ge 0  \nn\\
    &\quad  \mbox{ and } V_{ij,1}(x,\nu) \ge 0 \},
\end{align}
and
\begin{equation}
        \bar \SSS \triangleq \bar \SH \cap \SSS_\rms \cap ( \SR_1 \times \BBR^{\ell_\nu} ). 
\end{equation}
Proposition~1 in \cite{tan2021} implies that if $(x_0,\nu_0) \in \bar \SSS$ and for all $t\ge 0$, $(x(t),\nu(t)) \in \bar \SH$, then all $t\ge 0$, $(x(t),\nu(t)) \in \bar \SSS \subset \SSS_\rms$.
In this case, \ref{obj2}--\ref{obj4} are satisfied. 
Thus, we consider a candidate CBF whose zero-superlevel set is a subset of $\bar \SH$.
Specifically, let $\rho>0$, and consider $h \colon \BBR^{6n} \times \BBR^{\ell_\nu} \rightarrow \mathbb{R}$ defined by
\begin{align}
h(x,\nu) &\triangleq \mathrm{softmin}_{\rho} \Big ( R_{12,2}(x,\nu),\ldots,R_{1n,2}(x,\nu),\nn\\
&\qquad R_{23,2}(x,\nu),\ldots, R_{2n,2}(x,\nu),\ldots, R_{n-1n,2}(x,\nu),\nn\\
&\qquad V_{12,1}(x,\nu),\ldots,V_{1n,1}(x,\nu), V_{23,1}(x,\nu),\nn\\
&\qquad \ldots, V_{2n,1}(x,\nu), \ldots, V_{n-1n,1}(x,\nu),\nn\\
&\qquad Q_1(x,\nu),\ldots,Q_n(x,\nu) \Big ), \label{eq:h}
\end{align}
where $\mbox{softmin}_\rho : \BBR \times \cdots \times \BBR \to \BBR$ defined by 
\begin{equation}\label{eq:softmin}
\mbox{softmin}_\rho (z_1,\ldots,z_N) \triangleq -\frac{1}{\rho}\log\sum_{i=1}^Ne^{-\rho z_i},
\end{equation}
is the log-sum-exponential \textit{soft minimum}.

The soft minimum \eqref{eq:softmin} provides a lower bound on the minimum (e.g., \cite{rabiee2024c,rabiee2024b}). 
The zero-superlevel set of $h$ is 
\begin{equation}
\mathcal{H} \triangleq \{ (x,\nu) \in \mathrm{R}^{ 6n } \times \BBR^{\ell_\nu} \colon h(x,\nu) \geq 0 \}.
\end{equation}
Thus, $\SH$ is a subset of $\bar \SH$.
In fact, \cite[Prop. 2]{safari2024b} shows that as $\rho \to \infty$, $\SH \to \bar \SH$.
Thus, for sufficiently large $\rho>0$, $\SH$ approximates $\bar \SH$.

The next result is from \cite[Prop. 3]{rabiee2024b} and provides conditions such that $h$ is a relaxed CBF, that is, $h$ satisfies the CBF condition on 
\begin{equation}
    \mathcal{B} \triangleq \{ (x,\nu) \in \mathrm{bd} \ \mathcal{H} : L_\phi h(x,\nu) \leq 0 \},
\end{equation}
which is a subset of the boundary of $\SH$. 
\begin{proposition}
Assume that for all $(x,\nu) \in \SB$, $\frac{\partial h(x,\nu)}{\partial \nu} \ne 0$. 
Then, for all $(x,\nu) \in \SB$,
\begin{equation*}
    \sup_{\hat \mu\in \BBR^{\ell_\nu}} L_\phi h(x,\nu) + L_G h(x,\nu) \hat \mu \ge 0.
\end{equation*}
\end{proposition}

Since $h$ is a relaxed CBF, Nagumo's theorem implies that there is a control such that $\SH$ is forward invariant with respect to the dynamics \Cref{eq:cascade_1,eq:cascade_2}. 
However, $\SH$ is not necessarily a subset of $\SSS_\rms$. 
The next result follows from \cite[Prop. 3]{rabiee2024b} and is  useful because it shows that forward invariance of $\SH$ implies forward invariance of 
\begin{equation}
        \SSS \triangleq \SH \cap \SSS_\rms \cap ( \SR_1 \times \BBR^{\ell_\nu} ),
\end{equation}
which is a subset of $\SSS_\rms$. 

\begin{proposition}
Consider \Cref{eqn:State_space_n_satellites,eq:A,eq:B,eq:mu,eq:gamma} and \Cref{eqn:controller_state_space}, where $(x_0,\nu_0) \in \SSS$. 
Assume there exists $\bar t \in (0,\infty]$ such that for all $t \in [0,\bar t)$, $(x(t),\nu(t)) \in \SH$.
Then, for all $t \in [0,\bar t)$, $(x(t),\nu(t)) \in \SSS$.
\end{proposition}

\subsection{Optimal Control Subject to State and Input Constraints}

We now use the relaxed CBF $h$ to construct a constraint that guarantees \ref{obj2}--\ref{obj4}.
We also present a surrogate control for the control dynamics \eqref{eqn:controller_state_space} that is as close as possible to the desired surrogate control $\mu_\rmd$ while satisfying the constraint that guarantees that \ref{obj2}--\ref{obj4}.

Consider the constraint function $b \colon \BBR^{6n} \times \BBR^{\ell_\nu} \times \BBR^{\ell_\nu} \times \BBR \to \BBR$ given by 
\begin{align}
    b(x,\nu,\hat \mu, \hat \eta) &\triangleq L_\phi h( x,\nu) + L_G h(x,\nu) \hat \mu \nn\\
    &\qquad + \alpha ( h( x,\nu )) + \hat{\eta} h(x,\nu), \label{eq:b}
\end{align}
where $\hat \mu$ is the control variable; $\hat \eta$ is a slack variable; and $\alpha \colon \BBR \to \BBR$ is locally Lipschitz and nondecreasing such that $\alpha(0)=0$. 
Next, let $\gamma >0$ and consider the cost function %\SJ$ given by 
\begin{equation}
    \SJ(x,\nu,\hat \mu,\hat \eta) = \frac{1}{2} \| \hat \mu - \mu_\rmd(x,\nu) \|^2 + \frac{1}{2} \gamma \hat{\eta}^2. \label{eq:SJ}
\end{equation}
The objective is to find $(\hat \mu,\hat \eta)$ that minimizes the cost $\SJ$ subject to the relaxed CBF constraint $b(x,\nu,\hat \mu,\hat \eta) \ge 0$.

For each $(x,\nu) \in \BBR^{6n} \times \BBR^{\ell_\nu}$, the minimizer of $\SJ(x,\nu,\hat \mu,\hat \eta)$ subject to  $b(x,\nu,\hat \mu,\hat \eta) \ge 0$ can be derived from the first-order necessary conditions for optimality. 
For example, see \cite{rabiee2024b,safari2024b,ames2016,wieland2007,cortez2022}. 
This yields the control $\mu_* \colon \BBR^{6n} \times \BBR^{\ell_\nu} \to \BBR^{\ell_\nu}$ defined by
\begin{equation}    
\mu_*(x,\nu) = \mu_{\mathrm{d}} (x,\nu) + \lambda(x,\nu) L_G h(x,\nu)^{\mathrm{T}}, \label{eq:mu*}
\end{equation}
and the slack variable $\eta_* \colon \BBR^{6n} \times \BBR^{\ell_\nu} \to \BBR$
\begin{equation*}
    \eta_*(x,\nu) \triangleq \gamma^{-1} h(x,\nu) \lambda(x,\nu), %\label{eq:eta*}
\end{equation*}
where 
\begin{gather}
\lambda(x,\nu) \triangleq \begin{cases}
\dfrac{-\omega(x,\nu)}{ \| L_G h(x,\nu)^{\mathrm{T}}\|^2 + \gamma^{-1} h(x,\nu)^2 }, &\omega(x,\nu) <0,\\
        0, &\omega(x,\nu) \ge 0, \\
    \end{cases} \label{eq:lambda}\\
\omega(x,\nu) \triangleq b(x,\nu,\mu_\rmd(x,\nu),0). \label{eq:omega}
\end{gather}

The following result shows that $(\mu_*(x,\nu),\eta_*(x,\nu))$ is the unique global minimizer of $\SJ(x,\nu,\hat \mu,\hat \eta)$ subject to $b(x,\nu,\hat \mu,\hat \eta) \ge 0$.
The proof is similar to that of \cite[Theorem~1]{safari2024b}.

\begin{theorem} 
Assume that for all $(x,\nu) \in \SB$, $\frac{\partial h(x,\nu)}{\partial \nu} \ne 0$. 
Let $(x,\nu) \in \BBR^{6n} \times \BBR^{\ell_\nu}$, and let $(\hat \mu,\hat \eta) \in \BBR^{\ell_\nu} \times \BBR$ be such that $b(t,x,\hat \mu, \hat \eta) \ge 0$ and $( \hat \mu, \hat \eta) \ne (\mu_*(x,\nu), \eta_*(x,\nu))$. 
Then,
\begin{equation*}
    \SJ(x,\nu,\hat \mu,\hat \eta) >      \SJ(x,\nu,\mu_*(x,\nu),\eta_*(x,\nu)).
\end{equation*}
\end{theorem}

The following theorem is the main result on constraint satisfaction. 
It demonstrates that the control makes $\SSS$ forward invariant. 
This result follows from \cite[Corollary 3]{rabiee2024b}

\begin{theorem} 
Consider \Cref{eqn:State_space_n_satellites,eq:A,eq:B,eq:mu,eq:gamma}, where the control $\nu$ is given by \Cref{eqn:controller_state_space} with $\mu = \mu_*$, which is given by \cref{eq:mu*,eq:lambda,eq:omega}. 
Assume that for all $(x,\nu) \in \SB$, $\frac{\partial h(x,\nu)}{\partial \nu} \ne 0$, and assume that $h^\prime$ is locally Lipschitz. 
Then, for all $(x_0,\nu_0) \in \SSS$, the following statements hold: 
\begin{enumerate}
    \item There exists a maximum value $t_{\rm m} (x_0,\nu_0 ) \in (0 ,\infty]$ such that \Cref{eqn:State_space_n_satellites,eq:A,eq:B,eq:mu,eq:gamma,eqn:controller_state_space} with $\mu = \mu_*$ has a unique solution on $[0, t_{\rm m} (x_0,\nu_0 ))$.
    
    \item For all $t \in [0, t_{\rm m} (x_0,\nu_0 ))$, $(x(t),\nu(t)) \in \SSS \subset \SSS_\rms$.

%    \item Assume the maximum interval of existence and uniqueness is $t_{\rm m} (x_0,\nu_0 ) =\infty$. 
%    Then, \ref{obj2}--\ref{obj4} are satisfied. 

\end{enumerate}
\end{theorem}

Theorems 1 and 2 demonstrate that the control \Cref{eqn:controller_state_space,eq:mu*,eq:lambda,eq:omega} with $\mu = \mu_*$ satisfies \ref{obj2}--\ref{obj4}, and yields $\nu$ that is as close as possible to $\nu_\rmd$, which is designed to satisfy~\ref{obj1}.
\Cref{fig:block_diagram_cbf_emff} illustrates the implementation of the control \Cref{eqn:controller_state_space,eq:nu_d hat,eq:nu_d,eq:h,eqn:controller_state_space,eq:mu*,eq:lambda,eq:omega} with $\mu = \mu_*$.

\section{Formation Flying Simulation Results}
\label{sec:Formation Flying Simulation Results}

We present an example demonstrating EMFF of a system of 3 satellites modeled by (\ref{F_ij})--(\ref{accel_i}) with mass $m=15$~kg. 
They are equipped with orthogonal coils having $N=400$ turns and a cross-sectional area $A =0.1963$~m$^2$. 
The interaction frequencies are $\omega_{12}=200 \pi$~rad/s, $\omega_{13}=400 \pi$ rad/s, and $\omega_{23}=600 \pi$~rad/s. 
The resistance and inductance of each individual coil is $0.3673$~$\Omega$ and $0.12$~H. 
The relative position and velocity bounds are $\ubar{r}=1$~m and $\bar{v}=1$~m/s, and the power limit is $\bar{Q}=9 \times 10^6$~V$\cdot$A.

% \begin{example}
% \label{example_1}
This example demonstrates the no-collision position constraint and the power-limitation input constraint, while achieving formation. 
The initial conditions are $r_{1}(0) = \begin{bmatrix}
            1.2 &6.4 &8.5
        \end{bmatrix}^{\rm{T}}$~m, $r_{2}(0) = \begin{bmatrix}
            2.5  &7.5 &9  
        \end{bmatrix}^{\rm{T}}$~m, $r_3(0) = \begin{bmatrix}
            3.8  &8.6 &9.5
        \end{bmatrix}^{\rm{T}}$~m, and $v_1(0) = v_2(0) =v_3(0) = 0$~m/s.
        The desired relative positions are $d_{12} = \begin{bmatrix}
            1.1 &1.3 &0.5
        \end{bmatrix}^{\rm{T}}$~m, $d_{13} = \begin{bmatrix}
            2.2 &4.6 &1   
        \end{bmatrix}^{\rm{T}}$~m, and $d_{23} = \begin{bmatrix}
            1.1 &1.3 &0.5
        \end{bmatrix}^{\rm{T}}$~m. 
We let $a=0.7$, $\sigma=3$, $\rho=10$, $\alpha_0=\alpha_1=\alpha_v=5$, $\alpha=0.02$, and $\gamma=10^{40}$.
        
\begin{figure}[ht]
        \centering        \includegraphics[width=0.46\textwidth,clip=true,trim= 0in 0.05in 0in 0in]{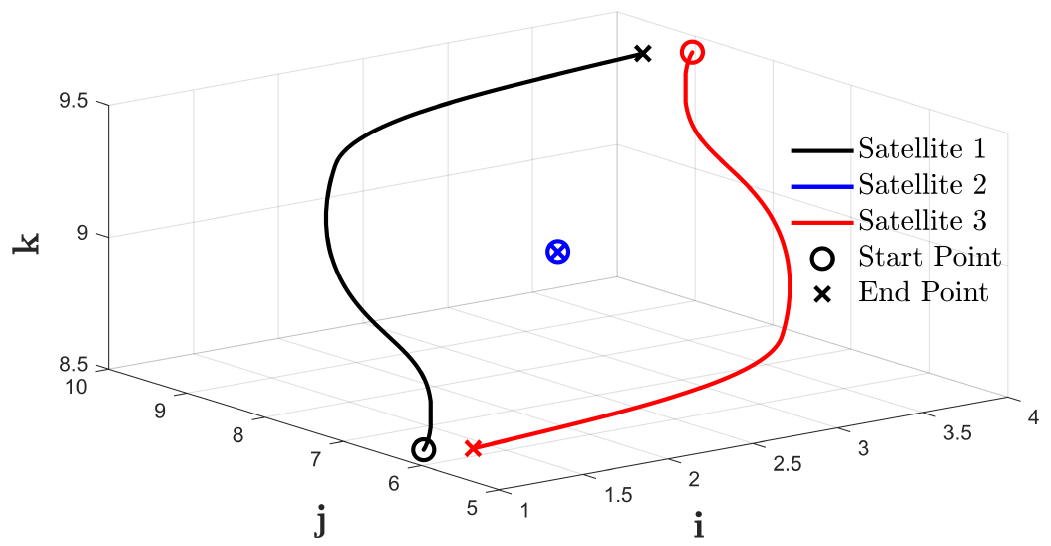}
        \centering
        \caption{Trajectories demonstrating collision avoidance.}
        \label{fig:3d_collision_avoidance}
    \end{figure}

\begin{figure}[ht]
        \centering        
        \includegraphics[width=0.46\textwidth,clip=true,trim= 0in 0.05in 0in 0in]{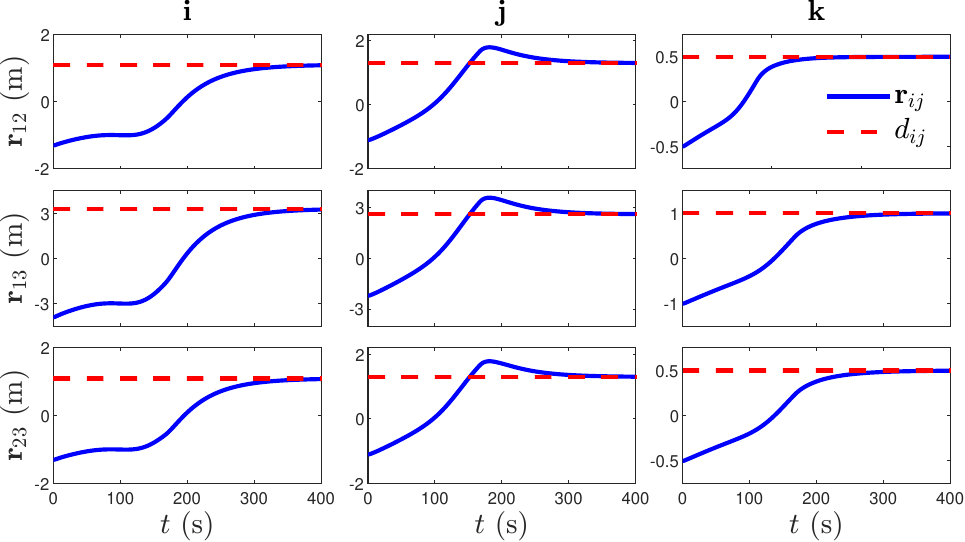}
        \caption{Relative positions $r_{12}$, $r_{13}$, and $r_{23}$ converge to $d_{12}$, $d_{13}$, and $d_{23}$ while avoiding collision.}
        \label{fig:position_power_plus_collision_avoidance}
    \end{figure}

\begin{figure}[ht]
        \centering        \includegraphics[width=0.46\textwidth,clip=true,trim= 0in 0.05in 0in 0in]{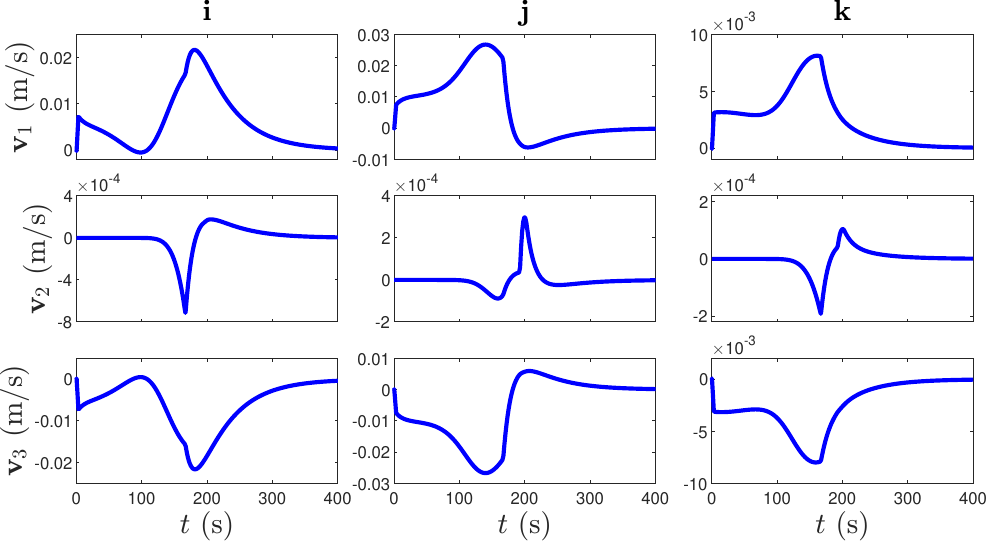}
        \caption{Velocities $v_{1}$, $v_{2}$, and $v_{3}$.}
       \label{fig:velocity_power_plus_collision_avoidance}
    \end{figure}

\begin{figure}[hbt!]
        \centering       \includegraphics[width=0.46\textwidth,clip=true,trim= 0in 0.05in 0in 0in]{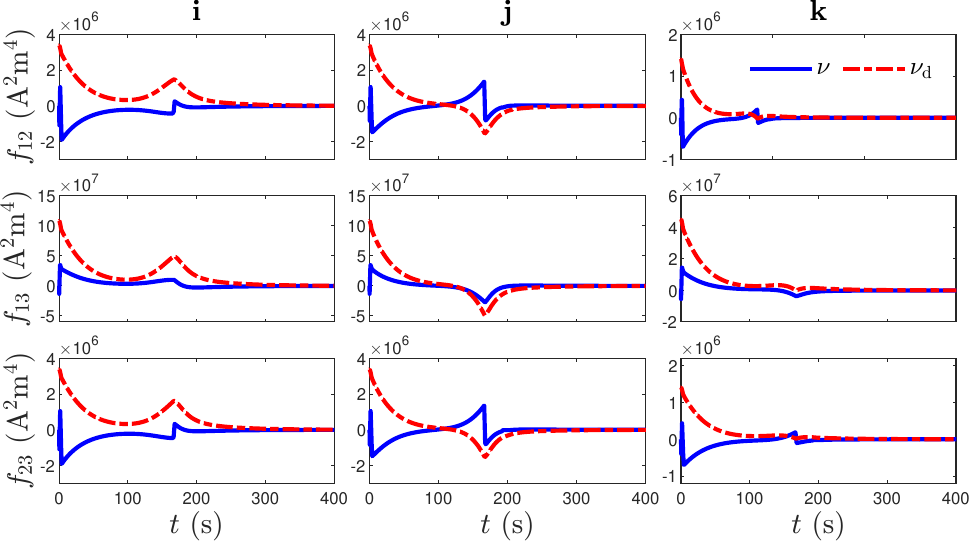}
        \caption{Intersatellite force functions.}
          \label{fig:force_power_plus_collision_avoidance}
    \end{figure}

\begin{figure}[hbt!]
        \centering     \includegraphics[width=0.46\textwidth,clip=true,trim= 0in 0.05in 0in 0in]{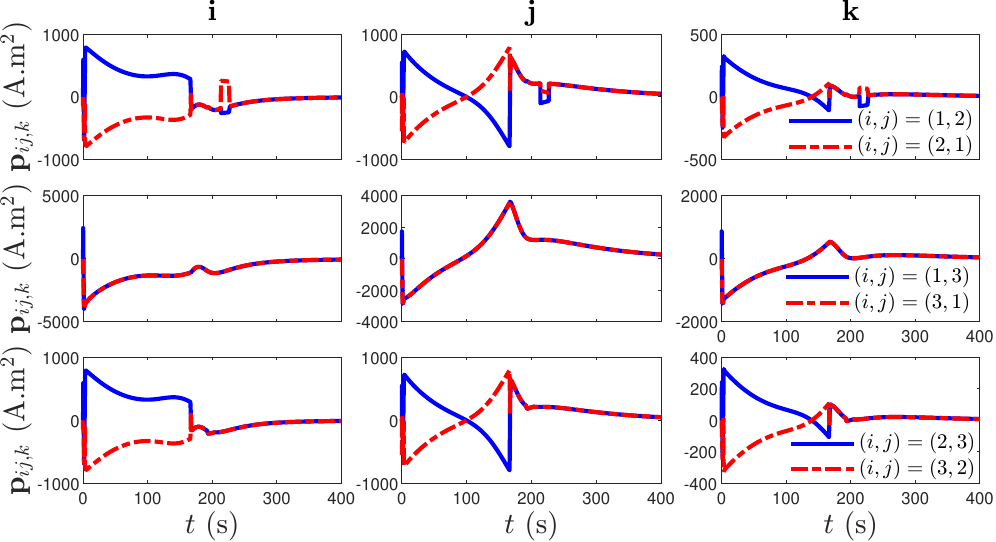}
        \caption{Amplitude controls.}
        \label{fig:amplitudes_power_plus_collision_avoidance}
    \end{figure}

\begin{figure}[hbt!]
        \centering      \includegraphics[width=0.46\textwidth,clip=true,trim= 0in 0.05in 0in 0in]{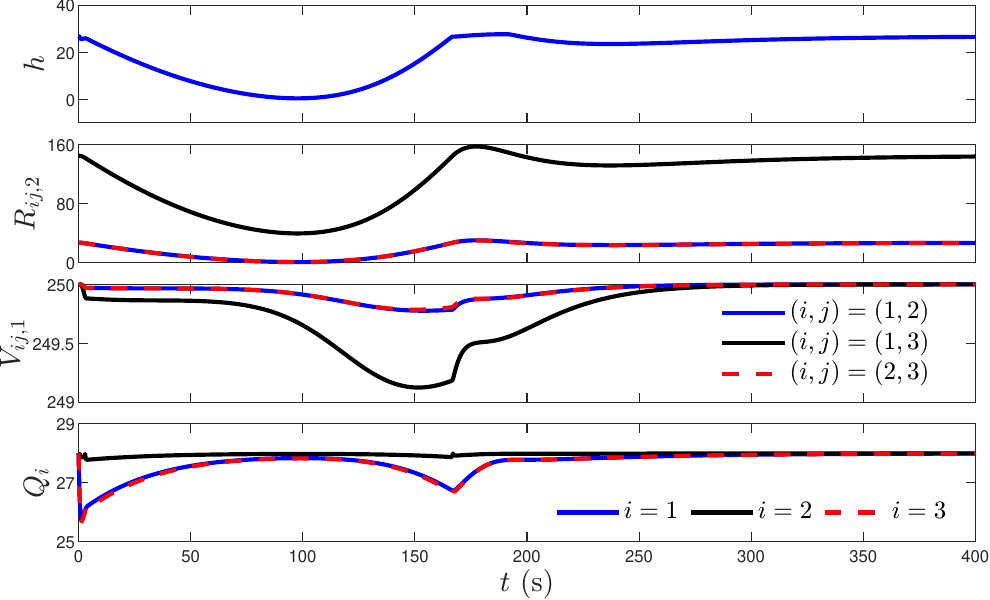}
        \centering
        \caption{Soft-minimum and its constituent barrier functions.} 
        \label{fig:barrier_functions_power_plus_collision_avoidance}
    \end{figure}

\Cref{fig:3d_collision_avoidance,fig:position_power_plus_collision_avoidance,fig:velocity_power_plus_collision_avoidance} shows the satellite trajectories and velocities.
The satellites avoid collision and achieve the desired formation.
Note that the straight paths from initial configuration to the desired formation would have resulted in collisions. 
\Cref{fig:force_power_plus_collision_avoidance} shows that the desired control $\nu_{\mathrm{d}}$ is modified in order to obtain the optimal-and-safe control $\nu$. 
\Cref{fig:barrier_functions_power_plus_collision_avoidance} shows $h$ and the the arguments of the soft minimum \eqref{eq:h} used to construct $h$.
The power-constraint barrier function $Q_3$ is the minimum argument of \eqref{eq:h} for $t\in[0.4,3.2]$~s and for $t \in [166.6,191.6]$~s. 
In contrast, the no-collision barrier functions $R_{12,2}$ and $R_{13,2}$ are the minimum argument of \eqref{eq:h} for $t\in(3.2,166.6)$~s. 
Thus, the modifications observed in \Cref{fig:force_power_plus_collision_avoidance} are made to satisfy both power and no-collision constraints. 
\Cref{fig:amplitudes_power_plus_collision_avoidance} shows the amplitude controls $\mathbf{p}_{ij,k}$ computed from $\nu$ using the \Cref{Section:Allocation}. 
% \exampletriangle
% \end{example}

    %
    \begin{figure}[hbt!]
        \centering        \includegraphics[width=0.46\textwidth,clip=true,trim= 0in 0.05in 0in 0in]{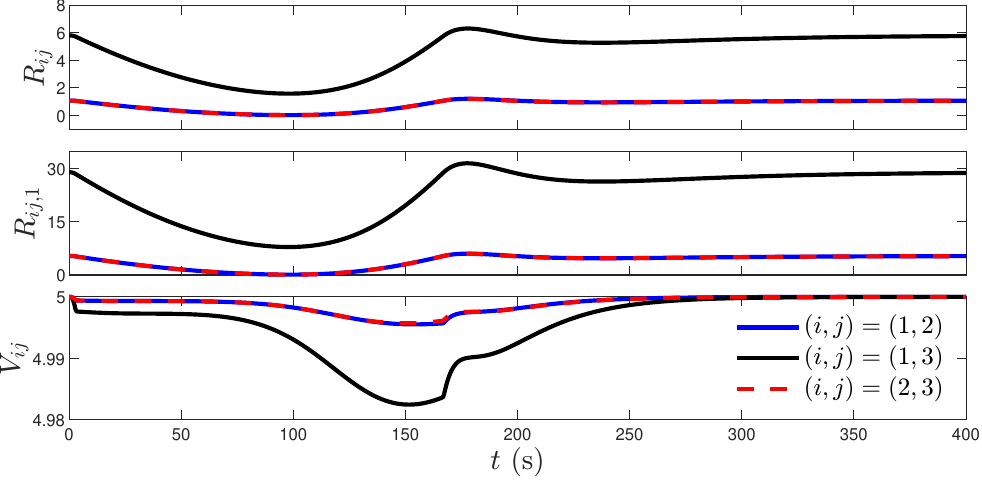}
        \centering
        \caption{Barrier functions $R_{ij}$, $R_{ij,1}$, and $V_{ij}$.} 
         \label{fig:R_ij2_barrier_functions}
    \end{figure}

 \bibliographystyle{elsarticle-num} 

 \bibliography{EMFF_CBF}

\end{document}